\documentclass[twocolumn, switch]{article} 

\usepackage{preprint}

\usepackage{enumitem}
\usepackage{amsmath, amsthm, amssymb, amsfonts}
\usepackage{bm}
\usepackage{amsmath}
\usepackage{graphicx}
\usepackage{subfigure}

\usepackage[algoruled]{algorithm2e}

\usepackage{setspace}
\usepackage{amssymb}
\usepackage{booktabs}
\usepackage{array}
\usepackage{color}
\usepackage{multirow}
\usepackage{multicol}
\usepackage{threeparttable}
\usepackage{url}
\usepackage[page]{appendix}

\usepackage[numbers,square]{natbib}
\usepackage[utf8]{inputenc}	
\usepackage[T1]{fontenc}	
\usepackage{xcolor}		
\usepackage[colorlinks = true,
            linkcolor = blue,
            urlcolor  = blue,
            citecolor = blue,
            anchorcolor = black]{hyperref}	
\usepackage{booktabs} 		
\usepackage{nicefrac}		
\usepackage{microtype}		
\usepackage{lineno}		
\usepackage{float}			

\usepackage{newfloat}
\DeclareFloatingEnvironment[name={Supplementary Figure}]{suppfigure}
\usepackage{sidecap}
\sidecaptionvpos{figure}{c}

\usepackage{titlesec}
\titlespacing\section{0pt}{12pt plus 3pt minus 3pt}{1pt plus 1pt minus 1pt}
\titlespacing\subsection{0pt}{10pt plus 3pt minus 3pt}{1pt plus 1pt minus 1pt}
\titlespacing\subsubsection{0pt}{8pt plus 3pt minus 3pt}{1pt plus 1pt minus 1pt}

\usepackage{graphics}
\usepackage{hyperref}
\usepackage{color}
\usepackage{multirow}
\usepackage{hhline}
\usepackage{subfigure}
\usepackage{lipsum}

\title{Debiased Cross-modal Matching for Content-based Micro-video Background Music Recommendation}

\usepackage{authblk}

\author{Jing Yi and Zhenzhong Chen*}
\affil{Wuhan University}

\begin{document}

\twocolumn[ 
  \begin{@twocolumnfalse} 
  
\maketitle

\begin{abstract}

Micro-video background music recommendation is a complicated task where the matching degree between videos and uploader-selected background music is a major issue. However, the selection of the user-generated content (UGC) is biased caused by knowledge limitations and historical preferences among music of each uploader. In this paper, we propose a Debiased Cross-Modal (DebCM) matching model to alleviate the influence of such selection bias.  
Specifically, we design a teacher-student network to utilize the matching of segments of music videos, which is professional-generated content (PGC) with specialized music-matching techniques, to better alleviate the bias caused by insufficient knowledge of users. The PGC data is captured by a teacher network to guide the matching of uploader-selected UGC data of the student network by KL-based knowledge transfer.  
In addition, uploaders' personal preferences of music genres are identified as confounders that spuriously correlate music embeddings and background music selections, resulting in the learned recommender system to over-recommend music from the majority groups. 
To resolve such confounders in the UGC data of the student network, backdoor adjustment is utilized to deconfound the spurious correlation between music embeddings and prediction scores. We further utilize  Monte Carlo (MC) estimator with batch-level average as the approximations to avoid integrating the entire confounder space calculated by the adjustment.
Extensive experiments on the TT-150k-genre dataset demonstrate the effectiveness of the proposed method towards the selection bias. The code is publicly available on: \url{https://github.com/jing-1/DebCM}.

\end{abstract}

\vspace{0.4cm}

  \end{@twocolumnfalse} 
] 

\newcommand\blfootnote[1]{%
\begingroup
\renewcommand\thefootnote{}\footnote{#1}%
\addtocounter{footnote}{-1}%
\endgroup
}

\section{INTRODUCTION}
\label{section1}

{\blfootnote{Corresponding author: Zhenzhong Chen, E-mail:zzchen@ieee.org}}Recent years have witnessed the prevalence of user-generated content platforms, where micro-videos have become a good multimedia carrier for people to record and share their daily lives. Micro-video platforms have attracted billions of users. Uploaders can record or make videos and select appropriate background music by themselves. However, the huge music pool makes manual selection time-consuming and labor-intensive, so it is necessary to recommend suitable background music for micro-videos \cite{liu2018background,jy}.  
 
Existing work on recommending background music to videos mainly focuses on designing models to achieve the matching of video and background music. Chen \textit{et al.} \cite{liu2018background} and Sasaki \textit{et al.} \cite{DBLP:conf/mmm/SasakiHOM15} relied on the emotion model in  \cite{thayer1990biopsychology} to divide music and videos into different emotional regions for matching. Chao \textit{et al.} \cite{chao2011tunesensor} leveraged the mood tags of music and videos to calculate the relatedness graph for recommending cross-media content. Li and Kumar \cite{li2019querybyvideo} utilized emotion tags as joint constraints to better align the embeddings of music and video modalities. However, the emotion tags were manually annotated, which takes lots of time and energy.
Establishing matched video-music pairs \cite{suris2018cross} facilities content-based background music recommendation, which could train matching models using latent factors without emotion tags.
Yi \textit{et al.} \cite{jy} have further established a large-scale video-music matching dataset from a micro-video sharing platform, which enables large amounts of UGC data to be used to train a better model.

\begin{figure}
\centering
\includegraphics[scale=0.55]{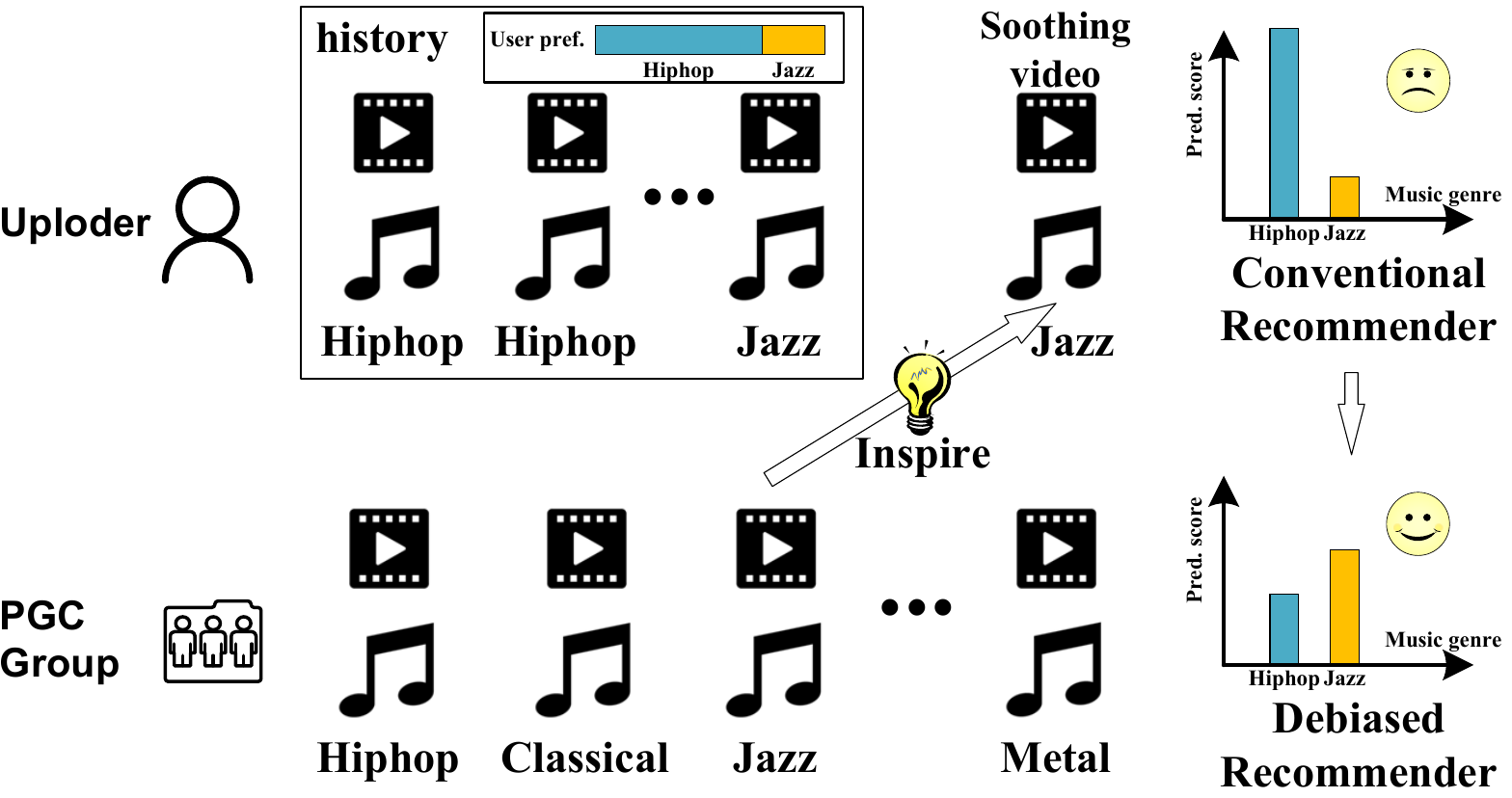}
\caption{Illustration of bias amplification brought about by traditional recommenders, and the debiased recommender we expect for more suitable matching by deconfounding the user historical genre preferences and giving more matching inspirations from the PGC.}
\label{FIG:illustration}
\end{figure}

However, the matching data are collected based on videos and user-selected background music, which results in selection bias from uploaders. First, the uploader has limited knowledge storage which hinders the selection of the best-matching music. Music video clips of professional-generated content could be used to inspire the matching of micro-videos and background music as illustrated in Fig. \ref{FIG:illustration}. For example, a soothing video which is a compilation of relaxing clips is mostly associated with light music in music videos that are composed by professional creation teams. Therefore, the co-occurrence of contents between videos and background music could inspire the recommenders to match Jazz music not rather noisy music, which may be missed by uploaders due to their limited music pool.
Second, uploaders have historical preferences among different music genres, which results in different selection probabilities for different music genres. For example, as shown in Fig. \ref{FIG:illustration}, users who prefer Hip-hop music are exposed to more Hip-hop music and are more likely to be attracted by Hip-hop music recommended by the recommender.  Such bias further causes the learned recommender to over-recommend the majority music genre, which will further intensify the exposure bias, thus resulting in the cocoon effect. Therefore,  the influence of such bias amplification should be alleviated.  
 
In consideration of the above biases, we design a Debiased Cross-Modal (DebCM) matching model for micro-video background music recommendations. Through the idea of knowledge distillation, we utilize a teacher network to capture the matching patterns of corresponding segments of PGC music videos, which embeds videos and music into a shared latent space with a variational encoder-decoder framework. A stable feature-learning inference model from the teacher network is further utilized to constrain and guide the learning of video and music embeddings of UGC data by the student network through KL divergence. In this way, the matching co-occurrence of PGC data could help the stable alignment of cross-modal embeddings so as to better alleviate the deviation caused by insufficient knowledge of uploaders.
Moreover, uploaders' historical preference for different music genres in UGC data is modeled as a confounder, which not only affects the music embeddings but also influences background music selections. To resolve the confounder, backdoor adjustment is utilized, where a Monte Carlo (MC) estimator with a batch-level average method is used to calculate the approximation of backdoor adjustment. 
In addition, to evaluate the debiased effect, we extend the TT-150k dataset \cite{jy} with attached music genres, where skewed test sets of intervention are sampled for testing. Specifically, we sample: 1) the videos of uploaders with diverse historical interaction among music genres for the test set to verify the effect of the student network on the deconfounder of UGC data, and therefore, whether it can cope with uploaders with rich and varied preferences and recommend suitable rather than biased background music could be evaluated; 2) the videos with the highest matching degree are sampled as the test set to verify the effect of the proposed teacher-student network on debiasing, and therefore, whether the most matching background music can be recommended for the video could be evaluated.

The main contributions can be summarized as follows:

\begin{itemize}[]
    \item We propose to utilize the matching of PGC data to guide the learning of UGC data so as to alleviate the selection bias caused by the limited knowledge storage of uploaders. Specifically, we design a teacher-student network with KL-based knowledge transfer to enable the PGC teacher network to constrain and guide the embedding learning of the UGC student network. Therefore, embeddings of more-matched music and videos could be aligned to better recommend more content-matching music to videos.
    \item Considering that the uploaders' historical preference among music genres are confounders, which affect both the learning of music embeddings and the uploaders' choice of background music, backdoor adjustment is used to deconfound the spurious correlation.  A Monte Carlo (MC) estimator with a batch-level average method is utilized to approximate the estimation of backdoor adjustment given computational efficiency and unbiasedness of MC by random sampling. Therefore, over-recommendation over the majority of historically-interacted music genres could be alleviated by the deconfounder.
    \item An extended dataset TT-150k-genre has been established to evaluate the performance with intervened test sets without the requirement of unbiased matching which are inaccessible.  Experiments demonstrate that good performance is achieved for debiasing with the intervened test set.  

\end{itemize}

The remainder of this article is organized as follows. Section \ref{section2} gives a related literature review of the work and Section \ref{section3} defines the task from a causal view. In Section \ref{section4}, we illustrate the proposed DebCM with details. Section \ref{section5} presents the experimental settings and analyzes the experimental results. Finally, Section \ref{section6} briefly summarizes the article.

\section{RELATED WORK} \label{section2}
\subsection{Audiovisual Cross-modal Matching}

Cross-modal matching aims to match queries and targets that come from two different modalities \cite{cross_modal_review}, which requires measuring the similarity of different modalities. Therefore, the alignment of two modalities is the essence of the problem. 

Existing work on audiovisual cross-modal matching could be classified into emotion tag-based and content-based methods. Emotion tag-based work uses the relatedness between visual tags and music mood tags \cite{chao2011tunesensor,li2019querybyvideo,9381371}. Moreover,  emotion model \cite{thayer1990biopsychology} has also been used to match video-music pairs \cite{liu2018background,DBLP:conf/mmm/SasakiHOM15}.
These emotion tags for the video, however, were manually annotated from crowdsourcing, which is high in labor and time costs.
Content-based methods utilize video and music contents of matched pairs to explore the matching patterns of cross-modal contents.
Canonical correlation analysis (CCA) based methods such as  \cite{music_video} were introduced to model the relationship between videos and music. Wu \textit{et al.} \cite{music_video} adopted lyrics as a middle media to connect music and image, and designed a set of lyric-based attributes for image representation. Zhang \textit{et al.} \cite{9320535} proposed a novel Adversarial-Metric Learning (AML) model for speakers' audio-visual matching.
Suris \textit{et al.} \cite{suris2018cross} further employed the visual features and audio features provided by Youtube-8M \cite{abu2016youtube-8m} to constrain the visual and audio embeddings of the same video as close as possible and {predicted} the corresponding label of the video.  CBVMR \cite{hong2018cbvmr} introduced a content-based retrieval model without any metadata like emotions, where inter-modal ranking-based matching loss and soft intra-modal structure loss have been proposed. Pretet \textit{et al.} \cite{tmm_bm_rec} focused on shorter segments of videos and music for training and inference, where structure-aware music segmentation and alignment of sequences have been proposed for recommendations for more professional videos.
Yi \textit{et al.} \cite{jy} proposed a cross-generation strategy for better aligning the latent embedding of music and micro-videos. A large-scale dataset, TT-150k, which is composed of about 3,000 music clips and 150k micro-videos, was established for evaluation. 
However, since the matching status between video-music pairs is extracted from uploader-selected background music for micro-videos, selection bias should be considered to better find the causality of video-music matching and recommend unbiased music, which is the main focus of this article.

Some pioneering methods have been proposed to consider the bias problem in visual-language cross-modal matching. Zhang \textit{et al.} \cite{zhang2021towards} proposed data debiasing and model debiasing strategies for the temporal sentence grounding in video (TSGV) task, where moment temporal distribution was considered to be the main cause of bias.
Wen \textit{et al.} \cite{wen2021debiased} pointed out that some visual question answering (VQA) models tended to output the answer that occurs frequently in the dataset and ignored the query images, where two unimodal bias detection modules were applied to explicitly recognize and remove the negative biases but remain the positive bias.
Huang \textit{et al.} \cite{huang2021learning} viewed the bias problem from a perspective of noisy correspondence, which refers to mismatch paired samples. The authors proposed to divide the data into clean and noisy partitions based on the memorization effect of neural networks and then rectify the correspondence via an adaptive prediction model in a co-teaching manner.

\subsection{Debiased Recommendation}
Because the data of user behavior are observational but not experimental, there exist biases due to some factors, such as users' selection of items and the system's exposure to items.  To some extent, the bias would be amplified since the recommender system would recommend more monotonous items according to the propensity in observational data \cite{chen2020bias}, which further damages users' experience. Therefore, alleviating the bias in recommender systems has become a new direction in the field of recommendations.  
Selection bias mainly comes from Missing Not At Random (MNAR) data since users tend to rate the items they are interested in and they rarely rate the items they are not interested in. The observed scores are not representative samples of all the scores, thus resulting in selection bias. Many studies have been dedicated to solving the problem mainly based on methods as follows.

Propensity score-based methods such as \cite{rec_treatment} are derived from statistics, which add the Inverse Propensity score (IPS) to the data for debiasing.  Schnabel \textit{et al.} \cite{rec_treatment} addressed selection bias using IPS, where the propensity score was assumed to be the probability of each data being observed. 
Lee \textit{et al.} \cite{ips_recsys} considered that although click behaviors observed in the recommendation system could reflect users' preferences to some extent, the missing click data did not necessarily mean negative feedback from users (positive-unlabeled problem).  The authors proposed an unbiased evaluator using IPS, where debiased methods could be properly evaluated.
Lee \textit{et al.} \cite{du} proposed a dual recommender learning framework that simultaneously eliminated the bias of clicked and unclicked data. Specifically, the proposed loss function adopted two propensity weighting
to effectively estimate the true positive and negative preferences
from clicked and unclicked data.

Causal graph-based methods such as \cite{scm} assume the inherent causal effect of the data and use the causal graph to depict the causal effect, which can work well if reasonable assumptions about the data-generation mechanism are made.  
Zhang \textit{et al.} \cite{scm} proposed a popularity-bias deconfounding and adjusting model, where item popularity could directly influence interaction probability because most people have a herd mentality. Do calculus was then used to remove the influence of the parent node of the confounder.
Wang \textit{et al.} \cite{scm1} scrutinized the cause-effect factors for bias amplification, identifying the main reason lay in the confounder effect of imbalanced item distribution on user representation and prediction score.
Wang \textit{et al.} \cite{scm2} formulated the recommendation models
as a causal graph that reflected the cause-effect factors in
recommendation, and addressed the clickbait issue by performing
counterfactual inference on the causal graph.  By estimating the click likelihood of a user in the
counterfactual world, the direct effect of
exposure features and the clickbait issue could be reduced.
 
Uniform data-based methods such as \cite{uni} utilize more reasonable experimental data (\textit{e.g.}, randomly investigate users' ratings on items), which could directly weaken bias.  Liu \textit{et al.} \cite{uni} introduced the uniform data to alleviate the bias through several knowledge distillation methods, where straight-forward sample-based, label-based, feature-based, and network structure-based methods were involved.
Chen \textit{et al.} \cite{chen2021autodebias} leveraged another (small) set of uniform data to optimize the debiasing parameters by solving the bi-level optimization problem with meta-learning.

Some other work focuses on fairness \cite{beutel2019fairness}, diversity \cite{pd-gan} and calibration \cite{calibration} for recommender systems, which induces different loss functions to increase the fairness and diversity of recommendation.

\section{Causal View of Background Music Recommendation} \label{section3}
In this section, we formally define the task of background music recommendation from a causal perspective. For UGC data, the uploader's historical preference is identified as a confounder that spuriously
correlates music embeddings and music selections. While for PGC data, professional teams make the uniform matching between videos and background music. To effectively utilize the PGC data, a Teacher-Student knowledge distillation framework could be utilized for a debiased recommendation. Moreover, a causal graph is established for UGC data, where we introduce
backdoor adjustment to block uploaders’ influence such that the
causal effects of music contents on music selections of videos can be properly estimated.

\subsection{Problem Definition}
We assume there are uploader-video-music triplets in the form of $\{( u, \mathbf{v}, \mathbf{m})\}$, where $\mathbf{v} \in \mathcal{V}$, $\mathbf{m} \in \mathcal{M}$ and $u \in \mathcal{U}$. Each music $\mathbf{m} \in \mathcal{M}$ is associated with a music feature m from its audio clip and a music genre $g$. Each video $\mathbf{v} \in \mathcal{V}$ is associated with a visual feature $\mathbf{v}$ extracted from the image sequence.
The mapping $\mathrm{f}: \mathcal{V} \times \mathcal{M} \rightarrow \{0, 1\}$ includes a set of triplets $\{(\mathbf{v}, \mathbf{m}, y)\}$, where $y=\mathrm{f}(\mathbf{v}, \mathbf{m})$ is the indicator that depicts whether music $\mathbf{m}$ is matched to a video $\mathbf{v}$ chosen by the uploader.
Besides the matching information of user-generated video-music, we introduce external information of PGC data, which include a set of matched triads $\{(\mathbf{v}^{t}, \mathbf{m}^{t}, y^{t})\}$ from professional music video clips.
Given a new video $\mathbf{v}$, our goal is to retrieve a list of music candidates $\mathcal{C}(\mathbf{m}) \subset \mathcal{M}$ where each music $\mathbf{m} \in \mathcal{C}(\mathbf{m})$ is a potentially suitable match for the video based solely on their content features, irrespective of uploader’s personal historical preference, such that selection bias that distracts users from locating suitable background music can be eliminated.

\subsection{Causal Graph Modeling}

To gain more intuition about the problem, we simplify a causal graph as illustrated in the User Deconfounder module of Fig. \ref{FIG:MODEL} by carefully considering the causal effects of user historical preference, music, video, and matching status. Specifically, we focus on user historical preference among music genres as the confounder, which affects both the music embeddings and the matching status. In particular: $\mathbf{u}$ represents the historical preference among music genres of the uploader $u$, $\mathbf{z}_m$ and $\mathbf{z}_v$ denotes the music embedding and video embedding learned from models inferred from music contents $\mathbf{m}$ and video contents $\mathbf{v}$, $y$ denotes the matching status of music and videos allocated by the uploader.

The edges in the graph describe the causal relations between variables. Specifically, $\mathbf{u} \rightarrow \mathbf{z}_m$ denotes the user historical preference among music genres affects the exposure and modeling of user embedding, $\mathbf{u} \rightarrow y$ represents the user historical preference affects the final selection of the matching status $y$, $\mathbf{z}_m \rightarrow y$ denotes the music embeddings learned from the model affects the matching status by calculating the similarity of embeddings of music and video.  Owing to the co-effects of user preference among music genres between music embedding and matching status, recommender models are affected by the confounder and thus suffer from a spurious correlation between the music embedding and the prediction score.

To eliminate the spurious correlation caused by the confounder, 
experimentally, intervention on the confounder which forces uploaders' interaction distribution among music genres to be controlled could achieve the purpose. However, it is almost not feasible to collect a certain number of such experimental data, since it is necessary to control the exposure content of the recommendation system and the user's independent choice, which will greatly damage the user's experience. To enable the estimation of the causal effect from observational biased UGC data, DebCM resorts to the causal solution: backdoor adjustment \cite{pearl2016causal}.
According to the causal theory, since $\mathbf{u}$ affects both $\mathbf{z}_m$ and $y$, $\mathbf{u}$ is a confounder between $\mathbf{z}_m$ and $y$, resulting in the spurious correlation when estimating the correlation between $\mathbf{z}_m$ and $y$. To dissolve the spurious correlation, we struggle to eliminate the edge between user historical preference $\mathbf{u}$ and music embedding $\mathbf{z}_m$ through backdoor adjustment. The details will be illustrated in Section \ref{stu_net} of deconfounded cross-modal model for the student network.

\section{METHODOLOGY} \label{section4}

\begin{figure*}
\centering
\includegraphics[scale=1.2]{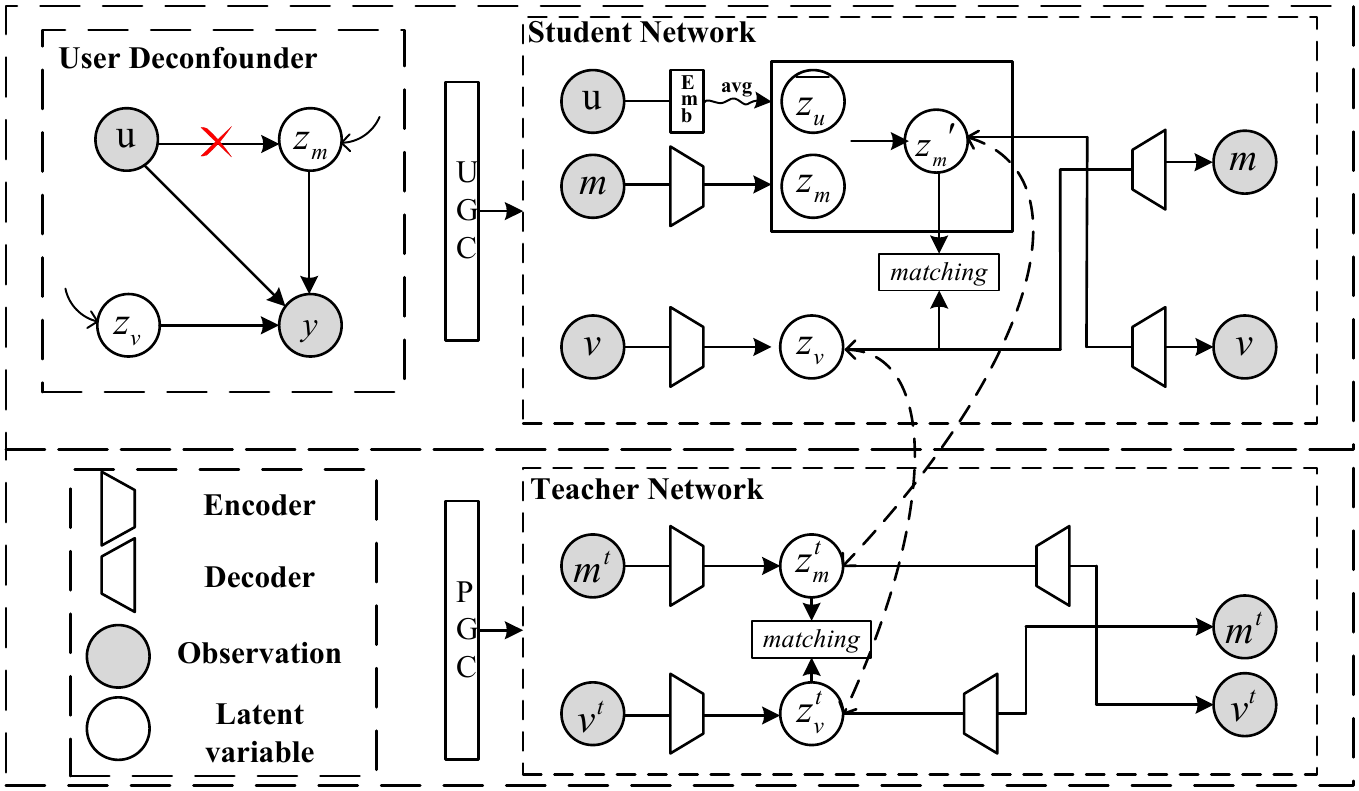}
\caption{The framework of our proposed DebCM for debiased background music recommendation of micro-videos. Specifically, the PGC teacher network is utilized to guide the embedding of the UGC student network through knowledge transfer. User preference among music genres, as a confounder, is further eliminated via a backdoor adjustment in the student network.}
\label{FIG:MODEL}
\end{figure*}

The overall framework of DebCM is shown in Fig. \ref{FIG:MODEL}. Specifically, DebCM aims to acquire professional knowledge from PGC data to alleviate the selection bias caused by limited knowledge of uploaders using a teacher-student knowledge transfer. To further debias the selection bias caused by uploaders' historical preference in the UGC data, we build up a causal graph and block the spurious relation caused by the confounder through backdoor adjustment in the student network. 
{Notations used in this article are summarized in Table \ref{TAB:notation}.}

\begin{table}[t]
\centering
\caption{{Notations used in our method.}}
\begin{tabular}{ll}
\toprule
  Notation& Description  \\ 
\midrule
$\mathbf{m} \in \mathcal{M}$& a music feature in music set $\mathcal{M}$ in UGC data\\
$\mathbf{v} \in \mathcal{V}$& a video feature in video set $\mathcal{V}$ in UGC data \\
$u \in \mathcal{U}$& a user $u$ in uploader set $\mathcal{U}$ in UGC data \\
$\mathbf{m}^t \in \mathcal{M}^t$& a music feature in music set $\mathcal{M}^t$ in PGC data\\
$\mathbf{v}^t \in \mathcal{V}^t$& a video feature in video set $\mathcal{V}^t$ in PGC data \\
$y$& indicator of the matching status\\
$\mathrm{f}$&  mapping of triads $\{(\mathbf{m}, \mathbf{v}, y)\}$ \\
$d$& the dimension of latent variables\\
$\mathbf{z}_m$&  the music latent variable of student network \\ 
$\mathbf{z}_v$&  the video latent variable of student network\\
$\mathbf{z}_m'$& the deconfounded music embedding of student network\\
$\mathbf{u}$& user historical distribution of music genres of user $u$\\
$\mathcal{U}_\mathbf{u}$& sampling space of $\mathbf{u}$\\
$\mathbf{z}_u$& user historical preference embedding of user $u$ \\
$\mathbf{z}^t_m$&  the music latent variable of teacher network \\ 
$\mathbf{z}^t_v$&  the video latent variable of teacher network\\
$\mathbf{z}^t_m{'}$&  the music latent variable inferred by teacher network \\ 
$\mathbf{z}^t_v{'}$&  the video latent variable inferred by teacher network\\
$p(\mathbf{z} \mid \cdot)$& true posterior of latent variables \\
$q(\mathbf{z} \mid \cdot)$& variational posterior of latent variables\\
$p(\cdot \mid \mathbf{z})$& conditional likelihood of latent variables\\
$p( \mathbf{z})$& prior of latent variables\\
\bottomrule
\end{tabular}
\label{TAB:notation}
\end{table}


In our framework, the teacher network captures the well-matched PGC data, whereas the UGC data is modeled with the student network. We utilize a general distillation function based on the potentially useful knowledge from the PGC data to improve the learning of the biased UGC data. Specifically, we follow CMVAE \cite{jy} where a cross-modal generative process is used to model the matching of the music and videos in the teacher network. 
The well-modeled teacher network is then utilized as a stable inference extractor to guide the embedding of music and videos in the student network. Furthermore, the student network considers user preference among music genres as the confounder and conducts the backdoor adjustment to eliminate the impact of the confounder. The cross-modal generative process in the student network is then integrated with the user deconfounder module.
The details of the proposed model are expounded in the following sections.

\subsection{Overview of CMVAE}
CMVAE \cite{jy} introduces a variational cross-generation strategy for background music recommendation for micro-videos.
In this paper, we adapt the basic framework of cross-generation in CMVAE to our background music recommendation problem for better aligning the latent embeddings of music and videos and more robust learning with a deep variational manner.

According to CMVAE, a shared $d$-dimensional Gaussian latent space $\mathbb{R}^{d}$ could be learned given the matched video-music pairs $\{(\mathbf{v}, \mathbf{m})\}$, where the matching degree can be regularized to the distance between the embeddings.
Cross-generation is proposed to achieve the alignment of video and music pairs
where the video feature $\mathbf{v}$ is generated by the matched music latent embedding $\mathbf{z}_{m}$, and vice versa to the music feature $\mathbf{m}$ and $\mathbf{z}_{v}$. Based on the cross-generation strategy, 
the generation process can be formulated as:
\begin{align}
p(\mathbf{v} \mid \mathbf{m})&=\mathbb{E}_{\mathbf{z}_{m} \sim p\left(\mathbf{z}_{m} \mid \mathbf{m}\right)}\left[p\left(\mathbf{v} \mid \mathbf{z}_{m}\right)\right] \label{eq:recon_1} \\
p(\mathbf{m} \mid \mathbf{v})&=\mathbb{E}_{\mathbf{z}_{v} \sim p\left(\mathbf{z}_{v} \mid \mathbf{v}\right)}\left[p\left(\mathbf{m} \mid \mathbf{z}_{v}\right)\right].
\end{align}

Variational inference is utilized to find variational posteriors of latent variables which are intractable, where the minimization of the KL-divergence is equivalent to the maximization of the Evidence Lower BOund (ELBO) \cite{kingma2013auto}. {The reconstruction part $\mathcal{L}_{\text{cross\_recon }}$ of the ELBO is as follows:}
\begin{equation}
\begin{aligned}
\mathcal{L}_{\text{cross\_recon }}&=\mathbb{E}_{\mathbf{z}_m \sim q\left(\mathbf{z}_m \mid \mathbf{m}\right)} [p(\mathbf{v} \mid \mathbf{z}_m)] \\
&+ \mathbb{E}_{\mathbf{z}_{v} \sim q\left(\mathbf{z}_v \mid \mathbf{v}\right)} [p(\mathbf{m} \mid \mathbf{z}_v)],
\end{aligned}
\label{eq:L_recon_loss}
\end{equation}
\noindent which aims to reconstruct the input features with cross-modal latent variables. It is implemented by mean square error (mse) losses between the observations and reconstructed inputs.

The KL-divergence part $\mathcal{L}_{{KL}}$ of the ELBO serves as a regularizer which constraints the variational posterior to be close to a standard Normal distribution $\mathcal{N}(\mathbf{0}, \mathbf{I}_{d})$. Therefore, only recommendation-relevant information could be encoded into the latent variable in consideration of robustness to noise.
It can be formulated as:
\begin{equation}
\begin{aligned}
\mathcal{L}_{\text{KL}} &=  \mathbb{E}_{\mathbf{z}_v \sim q(\mathbf{z}_v \mid \mathbf{v})}\left[KL\left\|q\left(\mathbf{z}_{v} \mid \mathbf{v}\right), p\left(\mathbf{z}_{v}\right)\right\|\right]  \\
&+ \mathbb{E}_{\mathbf{z}_m \sim q(\mathbf{z}_m \mid \mathbf{m})}\left[KL\left\|q\left(\mathbf{z}_{m} \mid \mathbf{m}\right), p\left(\mathbf{z}_{m}\right)\right\|\right], \\
\end{aligned}
\label{eq:kl}
\end{equation}
\noindent where $p\left(\mathbf{z}_{m}\right)$ and $p\left(\mathbf{z}_{v}\right)$ are the priors for the music and video latent variables, respectively. {For Gaussian variables, the KL-divergence has a analytical solution.} 

Besides the cross-generation losses, the matching loss is further defined as the expection of the generative distribution $p(y \mid \mathbf{z}_{m}, \mathbf{z}_{v})$, which could be calculated as:
\begin{equation}
\begin{aligned}
    \mathcal{L}_{\text{matching }} &= p(y \mid \mathbf{m}, \mathbf{v}) \\
    &=\mathbb{E}_{\mathbf{z}_{m} \sim q\left(\mathbf{z}_{m} \mid \mathbf{m}\right), \mathbf{z}_{v} \sim q\left(\mathbf{z}_{v} \mid \mathbf{v}\right)}\left[p\left(y \mid \mathbf{z}_{m}, \mathbf{z}_{v}\right)\right], \label{eq:matching_loss}
\end{aligned}
\end{equation}
where the matching probability $p(y \mid \mathbf{m}, \mathbf{v})$ depicts the matching degree of the given video-music pair. {Dot product of the latent embeddings of video and music is used to calculate the matching degree of a video-music pair.} 

\subsection{Cross-modal Model for Teacher Network}
In this work, we utilize cross-generation loss, kl-divergence loss and margin-based matching loss to train our teacher network as:
\begin{equation}
\begin{aligned}
    \mathcal{L}^t =& \mathbb{E}_{\mathbf{z}_m^t \sim q\left(\mathbf{z}_m^t \mid \mathbf{m}^t\right)} [p(\mathbf{v}^t \mid \mathbf{z}_m^t) + KL\left\|q\left(\mathbf{z}_{m}^t \mid \mathbf{m}^t\right), p\left(\mathbf{z}_{m}^t\right)\right\|] \\
    +& \mathbb{E}_{\mathbf{z}_v^t \sim q\left(\mathbf{z}_v^t \mid \mathbf{v}^t\right)} [p(\mathbf{m}^t \mid \mathbf{z}_v^t) + KL\left\|q\left(\mathbf{z}_{v}^t \mid \mathbf{v}^t\right), p\left(\mathbf{z}_{v}^t\right)\right\|] \\
    +& \mathbb{E}_{\mathbf{z}_{m}^t \sim q\left(\mathbf{z}_{m}^t \mid \mathbf{m}^t\right), \mathbf{z}_{v}^t \sim q\left(\mathbf{z}_{v}^t \mid \mathbf{v}^t\right)}\left[p\left(y^t \mid \mathbf{z}_{m}^t, \mathbf{z}_{v}^t\right)\right],
\end{aligned}
\end{equation}
where a robust feature extractor of the inference network could be obtained. The validation set is excluded for selecting the best teacher network. The details of the training procedure of the teacher network are summarized in Algorithm \ref{alg:alg2} for reference.

\subsection{Deconfounded Cross-modal Model for Student Network} \label{stu_net}

\subsubsection{Backdoor Adjustment}
According to Pearl's theory of backdoor adjustment \cite{pearl2016causal}, we should control the confounder such that the backdoor path could be blocked, which is achieved by the do-calculus on the music node $\mathbf{z}_m$ in the causal graph. $do(\mathbf{z}_{m})$ can be intuitively seen as cutting off the edge $\mathbf{u} \rightarrow  \mathbf{z}_{m}$ in the causal graph and blocking the effect of $\mathbf{u}$ on $\mathbf{z}_{m}$ by controlling $\mathbf{z}_{m}$, and thus, the target of DebCM could be formulated as: $P(y \mid  do(\mathbf{z}_{m}), \mathbf{z}_{v})$, which could be formalized as: 
\begin{subequations}
\label{eq:backdoor_adj}
\begin{align}\footnotesize
&P(y \mid do(\mathbf{z}_{m}), \mathbf{z}_{v} ) \notag \\
&=\int_{\mathcal{U}_\mathbf{u}} P(\mathbf{u} \mid d o(\mathbf{z}_m), \mathbf{z}_v) P(y \mid d o(\mathbf{z}_m),\mathbf{z}_v, \mathbf{u}) d \mathbf{u} \\
&=\int_{\mathcal{U}_\mathbf{u}} P(\mathbf{u}) P(y \mid \mathbf{z}_m, \mathbf{z}_v, \mathbf{u}) d \mathbf{u},
\end{align}
\end{subequations}
where $\mathbf{u}$ denotes the historical distribution among music genres of user $u$, and $\mathcal{U}_\mathbf{u}$ represents the total sample space of $\mathbf{u}$. Moreover, Eq. (\ref{eq:backdoor_adj}a) follows the law of total probability and Bayesian rule, and Eq. (\ref{eq:backdoor_adj}b) is based on the definition of do-operator in \cite{pearl2016causal}.

Intuitively, user preference among music genres $\mathbf{u}$ has extensive possible values in a specific dataset, \textit{i.e.}, users have various historical distributions over music genres. In DebCM, we use an approximation approach demonstrated as follows to fulfill the estimation.

\subsubsection{Backdoor Adjustment Approximation}
We re-formalize the integral in Eq. (\ref{eq:backdoor_adj}b) into an expectation form as:
\begin{equation}
\begin{aligned}
    &\int_{\mathcal{U}_\mathbf{u}}  P(\mathbf{u}) P(y \mid \mathbf{z}_m, \mathbf{z}_v, \mathbf{u}) d \mathbf{u} \\
    &=\mathbb{E}_{P(\mathbf{u})}[P(y \mid \mathbf{z}_m,\mathbf{z}_v, \mathbf{u})] 
    \approx P\left(\mathbf{z}_m, \mathbf{z}_v, \mathbb{E}_{P(\mathbf{u})} \mathbf{u}\right),
\end{aligned}    
\end{equation}
where the approximation is introduced by work in \cite{statistics} with a theoretical upper bound of the error. 
We estimate the expectation with a Monte Carlo method \cite{monte} by drawing samples from the space of user preference and calculating the expectation.
For the consideration of computational efficiency and variance reduction of the Monte Carlo estimator, we utilize the average of user preference among music genres in each batch to do the approximation.

\subsubsection{Backdoor Adjustment Estimator}
To facilitate the usage of DebCM, we design the operator to instantiate backdoor adjustment. Specifically, we infer the user preference embedding $\mathbf{z}_u$ of uploader $u$ from user historical distribution of interacted music genre $\mathbf{u}$ with an Embedding operation, where the expectation is calculated by statistics of the collected user space. 
To further embed the user preference among music genres, we instantiate a genre embedding as $E_g \in \mathbb{R}^{ d \times N_{g}}$, where $N_{g}$ is the number of music genres and $d$ is the dimension of genre embeddings. Then, we define $g \in range(1, N_g)$ as the $g$-th music genre, ${\mathcal{M}_u}$ as the set of collected music of uploader $u$, and $Genre(\mathbf{m})$ as the function that returns the music genre of music $\mathbf{m}$.
In this way, we could calculate the $g$-th music genres distribution of uploader $u$ as ${\mathbf{u}}_{g}=\sum_{\mathbf{m} \in \mathcal{M}_{u}} \mathbb{I}(Genre({\mathbf{m}})= g) / |\mathcal{M}_{u}|$. For example, we denote $\mathbf{u}$ as [0.5, 0.5, 0, 0] when we have 4 music genres, with half interacted music being the first genre and half being the second genre. 
Finally, we calculate the user preference embedding of batch-level average $\overline {\mathbf{z}_u}$ as $E_g \cdot \sum_{u \in \mathcal{U}_B} \mathbf{u} P( u)$ with $P(u)$ defined as $1/|\mathcal{U}_B|$ and $\mathcal{U}_B$ denotes the set of uploaders in a Batch during training, and then we concatenate it to music embedding $\mathbf{z}_m$ to gain deconfounded music embedding $\mathbf{z}_m'$ as:
\begin{equation}
    \mathbf{z}_m' = \mathbf{z}_m || \overline {\mathbf{z}_u}.
    \label{eq:concat}
\end{equation}

Then, cross-generation strategy and margin-based matching are conducted on $\mathbf{z}_m'$ and $\mathbf{z}_v$ to do the cross-modal alignment, where cross-generation loss, KL-divergence loss, and matching loss are calculated and added to get the loss of student network $\mathcal{L}^s$.

\subsection{Knowledge Transfer}
We propose to learn a stable inference extractor of the teacher network, and then, the embedding learning process of the student network could be guided by the knowledge transferred from the teacher network. Therefore, the causal features in the PGC data modeled by the teacher network could be utilized to correct the bias from the UGC data modeled by the student network. Specifically, a feature-based distillation is leveraged
to filter out the representative stable features from the teacher
network, which is then used to guide a student network to
mimic the feature inference of the teacher model. The teacher algorithm optimizes a deep auto-encoder model for feature
modeling and guides the learning of embeddings from the
student network through a KL-divergence manner, where distributions of latent variables in the student network could be guided and better aligned through the knowledge in the teacher network.
The final objective $\mathcal{L}^s{'}$ of the student network is:
\begin{equation}
    \mathcal{L}^s{'} = \mathcal{L}^s+\lambda_{v}^{\text {CausE }} KL\left\|\mathbf{z}_{v}^t{'},\mathbf{z}_{v}\right\|+\lambda_{m}^{\text {CausE }} KL\left\|\mathbf{z}_{m}^t{'},\mathbf{z}_{m}'\right\|,
\end{equation}
where $\mathbf{z}_{v}^t{'}$ and $\mathbf{z}_{v}$ are embedding of video from teacher and student network, $\mathbf{z}_{m}^t{'}$ and $\mathbf{z}_{m}'$ are embedding of music from teacher and student network of the UGC data. Here, we model each variable as a Gaussian distribution, where the KL divergence of two Gaussian variables have an analytic expression. The training steps of the student network guided by the teacher network are summarized in Algorithm \ref{alg:alg1}.

\subsection{Inference Strategy}

After training, the learned recommender model by $P(y \mid d o(\mathbf{z}_{m}), \mathbf{z}_{v})$ and inference model of $\mathbf{z}_{m}'$ and $\mathbf{z}_{v}$ could be used for debiased background music recommendation. Specifically, for a newly uploaded UGC $\mathbf{v}$, $\mathbf{z}_{v}$ could be inferred using $p\left(\mathbf{z}_{v} \mid \mathbf{v}\right)$ and music embeddings in the music pool could be inferred using $p\left(\mathbf{z}_{m} \mid \mathbf{m}\right)$. We utilize the global average of user historical interacted distributions among music genres in the training set to calculate the averaged user embedding  $\overline{\mathbf{z}_u}$, where the deconfounded music embedding ${\mathbf{z}_m'}$ can be obtained by Eq. (\ref{eq:concat}). Finally, the dot product is utilized to calculate the matching degree of video $\mathbf{v}$ and music in the music pool, the matching scores are then
ranked where the top-$K$ music is selected for recommendations.

To some extent, bias might be beneficial to exclude the music genres they dislike which could improve the satisfaction of recommended music for users.
For example, users might only like jazz music so they don't listen to the music in other genres. In this case, a non-debiased personalized background music recommender system could learn the correlation between user preference among music genres and the choice of the background music of users, which can recommend user-preferred music for the uploaded items. However, on the other hand, users also prefer to select more suitable background music for the uploaded music to facilitate the spread and the index of the video. In this way, dynamically disentangling the effects between the user preference among music and the true appropriate matching of the music and the video is a further research area of our work.

\begin{algorithm}[t]
\DontPrintSemicolon  \KwIn{A video-music matching dataset $\mathcal{D}^t=\{\mathcal{V}^t, \mathcal{M}^t, \mathrm{f}^t\}$;\\
 \quad \quad $\Theta^t$ indicates the model parameters.}

Randomly initialize $\Theta^t$.\;
 \While{not converged}{
  Randomly sample a batch $\hat{\mathcal{D}^t}$ from $\mathcal{D}^t$.\;
  Compute $\mu_{m}^t, \sigma_{m}^t$ via variational encoders.\;
  Add the KL-divergence to the loss.\;
  Sample $\mathbf{\epsilon} \sim \mathcal{N}\left(\mathbf{0}, \mathbf{I}\right)$ and compute $\mathbf{z}_{m}^t$ and $\mathbf{z}_{v}^t$ via the reparametrization trick.\;
  Add the cross reconstruction loss and matching loss to  $\mathcal{L}^t$.\;
  Compute the gradient of loss $\nabla_{\Theta^t} \mathcal{L}^t$.\;
  
  Update $\Theta^t$ by taking stochastic gradient steps.\;
}
\Return{$\Theta^t$}\;
\KwOut{TeacherN model trained on dataset $\mathcal{D}^t$.}
\caption{{TeacherN-SGD:} Training TeacherN with SGD.}
\label{alg:alg2}
\end{algorithm}

\begin{algorithm}[t]
\DontPrintSemicolon  \KwIn{A video-music matching dataset $\mathcal{D}=\{u, \mathcal{V}, \mathcal{M}, \mathrm{f}\}$;\\
 \quad \quad the video set $\mathcal{V}$ contains the visual feature $\mathbf{v}$;\\
 \quad \quad the music set $\mathcal{M}$ contains the audio feature $\mathbf{m}$ and music genre $g$;\\
 \quad \quad the mapping $\mathrm{f} = \{(\mathbf{v}, \mathbf{m}, y)\}$ with $y \in \{0, 1\}$;\\
 \quad \quad $\Theta$ indicates the model parameters.}

Randomly initialize $\Theta$.\;
 \While{not converged}{
  Randomly sample a batch $\hat{\mathcal{D}}$ from $\mathcal{D}$.\;
  \ForAll{$mod \in \{m, v \}$}{
    Compute $\mu_{mod}, \sigma_{mod}$ via variational encoders.\;
  }
  Add the KL-divergence to the loss.\;
  Inference $\mathbf{\mu}_{v}^t{'}, \mathbf{\sigma}_{v}^t{'}$ and $\mathbf{\mu}_{m}^t{'}, \mathbf{\sigma}_{m}^t{'}$ from teacher network.\;
  Sample $\mathbf{\epsilon} \sim \mathcal{N}\left(\mathbf{0}, \mathbf{I}\right)$ and compute $\mathbf{z}_m$ and $\mathbf{z}_v$ via the reparametrization trick.\;
  Concatenate the batch-average genre embedding $\overline {\mathbf{z}_u}$ to $\mathbf{z}_{m}$ to get $\mathbf{z}_{m}'$ for deconfounder.\;
  Add the cross reconstruction loss and matching loss.\;
  Add KL-divergence of T-S embeddings to the loss.\;
  Compute the gradient of loss $\nabla_{\Theta} \mathcal{L}^s{'}$.\;
  
  Update $\Theta$ by taking stochastic gradient steps.\;
}
\Return{$\Theta$}\;
\KwOut{DebCM model trained on dataset $\mathcal{D}$.}
\caption{{DebCM-SGD:} Training DebCM with SGD.}
\label{alg:alg1}
\end{algorithm}

\section{EXPERIMENTS} \label{section5}
In this section, we conduct extensive experiments based on the established TT-150k-genre dataset to evaluate our proposed model DebCM by investigating the following research questions:
\begin{itemize}[leftmargin=*]
\item How does DebCM perform for the intervened micro-video background music recommendation compared with the baseline methods? {Among them, non-debiased and debiased methods are included for comprehensive comparisons.}
\item {Since the user preference deconfounder
and the PGC knowledge distillation modules are the main designs in our proposed DebCM, we investigate how do the two components contribute to the performance as ablation studies.}
\item {A case study is conducted to visualize whether DecRS can produce more robust and satisfying recommendations when user interest drift happens.}
\end{itemize}
 
\subsection{Dataset}
\begin{table}[htbp]
\centering
\caption{Statistics of the established TT-150k-genre dataset.}
\label{TAB:DATA_STATISTIC}
\begin{tabular}{cccccc}
\toprule
\#Music & \#Video  &\multicolumn{2}{c}{avg $\pm$ std \#v/\#m} &\multicolumn{2}{c}{min/max \#v/\#m}  \\ 
 3,003 &  146,351 &\multicolumn{2}{c}{49  $\pm$ 57} &\multicolumn{2}{c}{3 / 219}    \\
\midrule
 \#hiphop& \#jazz& \#classical& \#reggae& \#pop& \#metal \\
 651& 665& 1,330& 311& 42& 4 \\
\bottomrule
\end{tabular}
\end{table}

TT-150k \cite{jy} has been established for content-based micro-video background music recommendation which composes of approximately 3,000 different background music clips associated with 150,000 micro-videos from different uploaders. We extend it by categorizing each music with a music genre using a pre-trained music genre classification model \cite{8530557} where we call the extended dataset as TT-150k-genre. The pre-trained classification model is based on music genre from the GTZAN music corpus \cite{gtzan} with multiple layers of LSTM Neural Nets, which is then utilized to extract the music genre of each music clip in the TT-150k. 
Through getting the music genres, user preference among music, which is the confounder that exists in micro-video background music recommendation could be obtained by calculating her or his historical interactions among music genres. The statistics of the dataset are in Table \ref{TAB:DATA_STATISTIC}.

\textit{\textbf{Intervention among Test Set:}}
Since the dataset is generated by uploaders themselves, rather than experimental, it contains a lot of bias. Measurements on the biased test set are unlikely to be reliable.  However, it is time-consuming and difficult to collect a completely unbiased, content-based music-video matching data set, and therefore, we use a non-random sampling to produce an unbiased-like test set as the intervened test set. Followed by \cite{liang2016causal}, we sample two settings of the skewed test set to evaluate the effectiveness of our proposed user deconfounder module and PGC knowledge transfer network, respectively. Specifically, to evaluate the deconfounded ability of uploader's historical preference among music genres, we sample the videos with uploaders whose historical interactions among music genres are the most diverse as our skewed test set $\mathcal{T}_\text{diverse}$, where the diversity degree is calculated by the entropy of the probability among different music genres of uploader's historical interactions. In this way, the quality is assessed on videos with interest-varied uploaders, and therefore, the ability to alleviate the confounder effect which will cause the recommender to recall items with majority music genres of uploaders' historical preference could be well evaluated. On the other hand, to evaluate the effectiveness of PGC knowledge transfer, which aims to recommend the most-matching background music to the videos, we construct an intervened test set $\mathcal{T}_\text{matching}$ by sampling the videos that match the background music best. The matching degree is calculated by two factors: 1) the popularity of the videos; 2) the matching scores calculated by CMT \cite{di2021video}, where music-video rhythmic relations are utilized for referring to the matching degree of the music and the video. In this way, well-matched videos are sampled for evaluating the effectiveness of recommending the most-matching background music.

\subsection{Methods for Comparisons}
To demonstrate the effectiveness of the proposed DebCM, we draw comparisons with several state-of-the-art baselines, where debiased methods considering bias problems and non-debiased methods are both included. Moreover, all methods are tested with two settings: 1) using only UGC data and 2) using both UGC and PGC data. For non-debiased methods, we utilize the PGC data as the training data straightforwardly. Label-based \cite{uni} knowledge distillation is used to employ the PGC data for debiased methods because debiased methods need auxiliary genre data during training the UGC network which is not applicable for PGC data. More specifically, 
we have obtained a teacher model trained on PGC data, and then use it to predict all the UGC samples, where the imputed labels
are combined with the original labels through a weighting parameter to train a more unbiased student model on UGC data. The baselines are listed as follows:

\begin{itemize}[leftmargin=*]
    \item \textbf{CEVAR} \cite{suris2018cross}: It is a matching-based method that applies a cosine similarity loss to constrain the matched video and music embeddings to be close. A label prediction loss of the video is also attached where we dismissed it for applicable comparisons. 
    \item \textbf{CMVBR} \cite{hong2018cbvmr}: CMVBR leverages a ranking-based matching loss to align inter-modal embeddings of music and video pairs, where an intra-modal loss is applied to constrain similarity within one modality.
    \item \textbf{DSCMR} \cite{zhen2019deep}: It introduces a weight-sharing strategy for the last layer of the two-branch framework, and therefore, the cross-modal difference could be reduced and a matching-loss in the common latent space is utilized for matching.
    \item \textbf{UWML} \cite{wei2020universal}: It is a margin-based metric learning framework to constrain the closeness of matched pairs to unmatched pairs. Moreover, different weights are given to different samples according to the difficulty of matching.
    \item \textbf{Dual-VAE} \cite{shen2018deconvolutional}: Dual-VAE is a generative-based retrieval model, where question-to-question and
    answer-to-answer reconstruction loss are utilized for alignment of the matched pairs.
    \item \textbf{CMVAE} \cite{jy}: It introduces a PoE fusing module to integrate the visual and textual modalities in micro-videos. Moreover, a cross-generation strategy is utilized to further align the latent space of music and videos. It is the backbone of our proposed model where textual modality is dismissed for the lack of textual descriptions of PGC data.
\end{itemize}

We also compare our proposed debiased cross-modal background music recommender with the state-of-the-art debiased models in recommender systems:
\begin{itemize}[leftmargin=*]
    \item \textbf{IPS} \cite{ips_recsys}:  IPS is a statistics-based method in causal recommendation. Here we use user preference among music genres as
    the propensity of an uploader to down-weight the music in the majority genre group during debiased training.
    \item \textbf{DU} \cite{du}: It is an extension of IPS, which introduces a dual recommender learning framework that simultaneously eliminates the bias of clicked and unclicked data.
    \item \textbf{Calibration} \cite{calibration}: It proposes a calibration metric to keep the proportion of different areas of interest in the recommendation list to be the same as user historical interests for alliterating the over-recommend phenomenon, where post-processing toward the output of the recommender is utilized to minimize the metric.
    \item \textbf{DecRS} \cite{scm1}: DecRS is a deconfounded method that considers the user preference among item categories as a confounder and utilizes backdoor adjustment to alleviate the amplification of the bias problem, where the global-average estimation of the confounder has been utilized.

\end{itemize}
The above methods are model-agnostic, where we use our backbone and apply the debiased module for fair comparisons.

\subsection{Evaluation Metrics}
Two standard evaluation metrics for retrieving: Recall@$K$ and NDCG@$K$ \cite{wei2020universal} are leveraged to evaluate the model performance, which calculates the hit-ratios of ground-truth music on top-$K$ retrieved music. According to \cite{jy}, the long-tail characteristic of the popularity of different music clips is severe, which leads to a systematic bias that favors the popular music clips when weighing the match equally for different music clips. To alleviate the problem, following \cite{rec_treatment,jy},
music clips for evaluation are weighted by the inverse of their
popularity levels. Therefore, the Recall@$K$ and NDCG@$K$ we adopt in this paper are calculated as:
\begin{align}
    \text{Recall}@K &=\sum_{v \in \mathcal{V}^{t e} } \lambda_v \cdot \#\text {Hits }_{v} @ K \label{eq:recall}\\
    \text{NDCG}_{u} @ K&= \sum_{v \in \mathcal{V}^{t e} } {\sum_{i=1}^{K} \lambda_v \cdot \frac{\log _{2}2 \cdot \left({2^{r_{i}}-1}\right)}{\log _{2}(i+1)}} \\
    \lambda_v &= \frac{1 / P(v \rightarrow m)}{\sum_{i \in \mathcal{V}^{t e} } 1/P(i \rightarrow m)} \label{eq:weight},
\end{align}
where $\mathcal{V}^{t e}$ represents the set of test videos.
Recall@$k$ calculates the hit ratio of the tested models. Normalized Discounted Cumulative Gain (NDCG) assigns different importance to different ranks as $r_{i}$ represents the relevance degree of the item at position $i$ and it is assigned as binary (\textit{i.e.,} 0 or 1) in implicit recommendation scenario with ${\log2}$ to be a normalizer.
$\lambda_v$ denotes the weight of the video $v$ calculated by Eq. (\ref{eq:weight}). Specifically, $1 / P(v \rightarrow m)$ is assumed to re-weight the popularity bias of the music clips, which is calculated as the reciprocal of the popularity of the music clip that the target video v used as background music. Therefore, the popularity-debiased hit-ratio could be obtained with Recall@$K$ and NDCG@$K$ used in this paper.

\subsection{Parameter Settings}
Our DebCM model is implemented in Pytorch. The dimension of embeddings $d$ is fixed to 128, Adam \cite{duchi2011adaptive} is applied as the optimizer and the batch size is set to 1,024 for all models, empirically. For the proposed DebCM and all baselines, we adopt a grid search to select the hyperparameters based on evaluation metrics on the validation set.
The learning rate is search in \{0.0001, 0.005, 0.001, 0.05,
0.01\}, the L2 normalization coefficient is tuned amongst \{0.001, 0.01, 0.1, 0\}, and the dropout rate is selected in \{0.1, 0.2, 0.3, 0.4, 0.5\}.
Two fully-connected layers $[F  \rightarrow d \rightarrow F]$ with $F$ to
be the dimension of different features are applied as the encoder and decoder for the music and video networks of the cross-modal matching model. We further select the parameters of bi-directional margin-based matching loss by searching, and specifically, the weight of the music-to-video matching loss is set to 1, the margin of the matching loss is set to 0.05, and 40 most hard-to-match negative samples in the ranking-based matching loss are the chosen during training. By tuning, we set the weight of the L2 norm to 0.001, the weight of the reconstruction loss to 1, the dropout rate to 0.2,
and the learning rate to 0.001.
For the teacher distillation, we set the weight of the teacher guidance loss to be 40 by tuning.

\subsection{Performance Comparison with Baselines}

\begin{table*}[]
\centering
\caption{Comparison between the proposed DebCM and various baselines using only UGC data and with both UGC and PGC data.}
\label{TAB:result_with_different_methods}
\resizebox{\textwidth}{54mm}{
\begin{tabular}{l|l|cccc|cccc}
\toprule
&&\multicolumn{4}{c|}{\textbf{$D_s$}} & \multicolumn{4}{c}{\textbf{$D_s \cup D_t$}}\\
&&Recall@15 & NDCG@15 & Recall@25 &NDCG@25& Recall@15 & NDCG@15 & Recall@25 &NDCG@25 \\ 
\midrule
\multicolumn{10}{c}{$\mathcal{T}_\text{diverse}$} \\\midrule
&DebCM& \textbf{0.2887} & \textbf{0.0990} & \textbf{0.4047} & \textbf{0.1164} & \textbf{0.3093  } & \textbf{0.0999} & \textbf{0.4537} & \textbf{0.1176} 
\\   
\midrule
Non-Deb
&CEVAR \cite{suris2018cross}& {0.1433} & {0.0560} & {0.2437} & {0.0788} & {0.1415} & {0.0537} & {0.2409} & {0.0764} 
\\
&CMVBR \cite{hong2018cbvmr}& 0.1507& 0.0565& 0.2584& 0.0810& 0.1530& 0.0586& {0.2825}& 0.0875
\\
&DSCMR \cite{zhen2019deep}& 0.1626& 0.0568& 0.2955& 0.0869& {0.1763}& {0.0634}& 0.2868&  0.0888
\\
&UWML \cite{wei2020universal}& 0.1707& {0.0719}& 0.2800& {0.0967}& 0.1666& 0.0591& 0.2790&  0.0847\\
&Dual-VAE \cite{shen2018deconvolutional}& 0.1936& 0.0803& 0.2940& 0.1031& 0.1834& 0.0752& 0.2713& {0.0953}
\\
&CMVAE \cite{jy}& \underline{0.2452}& \underline{0.0941}& \underline{0.2945}& \underline{0.1049}& \underline{0.1961}& \underline{0.0821}& \underline{0.2880}& \underline{0.1019}
\\
\midrule
Deb& DebCM-IPS \cite{ips_recsys}& 0.1890& 0.0741& 0.3026& 0.1000& 0.1689& 0.0597& {0.2748}& {0.0946}
\\
&DebCM-DU \cite{du}& 0.1975& 0.0772& 0.3046& 0.1017& {0.2118}& 0.0875& 0.3178& 0.1004
\\
&DebCM-Calibration \cite{calibration}& 0.2071& 0.0802& 0.3174& 0.1029& \underline{0.2119}& \underline{0.0896}& {0.3188}& 0.1095
\\
&DebCM-DecRS \cite{scm1}& \underline{0.2392}& \underline{0.0818}& \underline{0.3606}& \underline{0.1045}& {0.2078}& {0.0831}& \underline{0.3412}& \underline{0.1135} \\

\midrule

\multicolumn{10}{c}{$\mathcal{T}_\text{matching}$} \\\midrule
&DebCM& \textbf{0.2529} & \textbf{0.0882} & \textbf{0.3865} & \textbf{0.1114} & \textbf{ 0.3293} & \textbf{0.1065} & \textbf{0.4633} & \textbf{0.1236} 
\\   
\midrule
Non-Deb
&CEVAR \cite{suris2018cross}& {0.1670} & {0.0731} & {0.2722 } & {0.0970} & {0.1415} & {0.0537} & {0.2409} & {0.0764} 
\\
&CMVBR \cite{hong2018cbvmr}& 0.1630& 0.0723& 0.2662& 0.0931& 0.1912& 0.0775& {0.2912}& 0.1038
\\
&DSCMR \cite{zhen2019deep}& 0.1503 & 0.0627& 0.2541& 0.0861& {0.1692}& {0.0638}& 0.2657 & 0.0857
\\
&UWML \cite{wei2020universal}& 0.1583 & {0.0567}& 0.2628& {0.0805}& 0.1569 & 0.0655& 0.2597 &0.0890  \\
&Dual-VAE \cite{shen2018deconvolutional}& 0.2150 &0.0925 & 0.3165& 0.1156& 0.1782& 0.0764& 0.2600 & 0.0950
\\
&CMVAE \cite{jy}& \underline{0.2390}& \underline{0.1036}& \underline{0.3345 }& \underline{0.1253}& \underline{0.2328}& \underline{0.1004}& \underline{0.3328}& \underline{0.1232}
\\
\midrule
Deb& DebCM-IPS \cite{ips_recsys}& 0.1806 &0.0696& 0.3083 &0.0986 & 0.1734& 0.0756& 0.2873& {0.1003}
\\
&DebCM-DU \cite{du}&0.1912   & 0.0785& 0.3255& 0.1012& {0.2394}& 0.0796& \underline{0.3512}& \underline{0.1202}
\\
&DebCM-Calibration \cite{calibration}& 0.2428& 0.0811& 0.3478& {0.1014}& \underline{0.2436}& \underline{0.0863}& {0.3484}& {0.1138}
\\
&DebCM-DecRS \cite{scm1}& \underline{0.2457 }& \underline{0.0821}& \underline{0.3617 }& \underline{0.1017}& {0.2164 }& {0.0849}& {0.3369}& {0.1121} \\



\bottomrule
\end{tabular}
}
\end{table*}

The overall comparison results are summarized in Table \ref{TAB:result_with_different_methods}.
From the table, we could see that the majority of non-debiased methods yield inferior results than debiased methods on both $\mathcal{T}_\text{diverse}$ and $\mathcal{T}_\text{matching}$, where intervention is performed on the testing data.  Among the non-debiased methods, generative-based methods (Dual-VAE and CMVAE) achieve improvements over matching-based methods, which demonstrates the effectiveness of generative-based models and variational-based training, which could be more robust to noise and disturbances. While on the intervened test set, the non-debiased methods that are based on strong inductive loss in an unbiased hypothesis, some relational effect, not the causal effect, could be learned that is harmful to the intervened test set so as to make the result not ideal.  In addition, simply adding PGC data to UGC data for the training set cannot achieve good results. The possible reason is that PGC data is more diverse, which makes training more difficult and contains more uncertainty.  

For debiased methods,  IPS performs worse than the vanilla backbone model (CMVAE) on $\mathcal{T}_\text{diverse}$ settings, which suggests that re-weighting the samples is not enough to alleviate the bias, especially when there exist not-really-matching music-video pairs caused by the selection bias of limited knowledge of uploaders. 
DU adds bidirectional constraint on IPS, thus gaining some additional improvement. Calibration re-ranks the matching sequence which could perform well on our skewed test set by considering the deconfounder degree and matching degree.
However, the above methods that don't carefully consider user preference among music genres as confounders and truly content-matching of music-videos could not achieve splendid results.
Structure causal modeling (SCM) -based approach DecRS works well for causal reasoning by establishing true causal relationships and removing harmful associations (the influence of the uploader's historical preference for music embeddings and music choice).  
For the use of PGC data, a label-based knowledge distillation method is utilized.  It can be seen from the results that the ideal guidance cannot be obtained by using PGC data in this way, because the prediction scores of the PGC teacher network directly change the correctness of UGC data with strong labels, while in fact, the teacher model inevitably contains errors brought by model, data and other factors.  

DebCM proposed in this work further optimizes the backdoor approximation over DecRS in the student network, which replaces the global average with the batch-level average to make the model conform to the unbiased substitution of Monte Carlo sampling.  From a macro point of view, the batch-level average introduces more randomness into the training, so that the model can better deal with data deviation. Therefore, our proposed DebCM could outperform other methods on  $\mathcal{T}_\text{diverse}$, which aims to evaluate the effectiveness of deconfounder ability caused by the uploader's historical preference. As a by-product, the deconfounder module in the student network could, on the other hand, improve true-matching between videos and music by alleviating the over-recommendation of majority groups of music.  Furthermore, to employ PGC data to further eliminate the UGC selection bias caused by the uploader's personal limitation in the data, we put forward PGC training teachers to guide the UGC training of the student network. As we could see from the results of using \textbf{$D_s \cup D_t$} on both settings of our proposed DebCM, the learning process of hidden variables in the student network is further constrained and guided by the teacher network, so as to get better performance.


\subsection{Ablation Study of DebCM}
In this section, we investigate the effectiveness of different components in DebCM. In detail, as the user preference deconfounder plays a vital role in DebCM, we change different ratios of intervention on music genre in training and test sets to investigate the effect of the deconfounder module. Moreover, we explore the effectiveness of PGC knowledge distillation by changing different weights of the PGC teacher network in the proposed model.

\subsubsection{The Effectiveness of User Preference Deconfounder}

\begin{figure}
\centering
\subfigure[Recall@15]{
\centering
\includegraphics[scale=0.29]{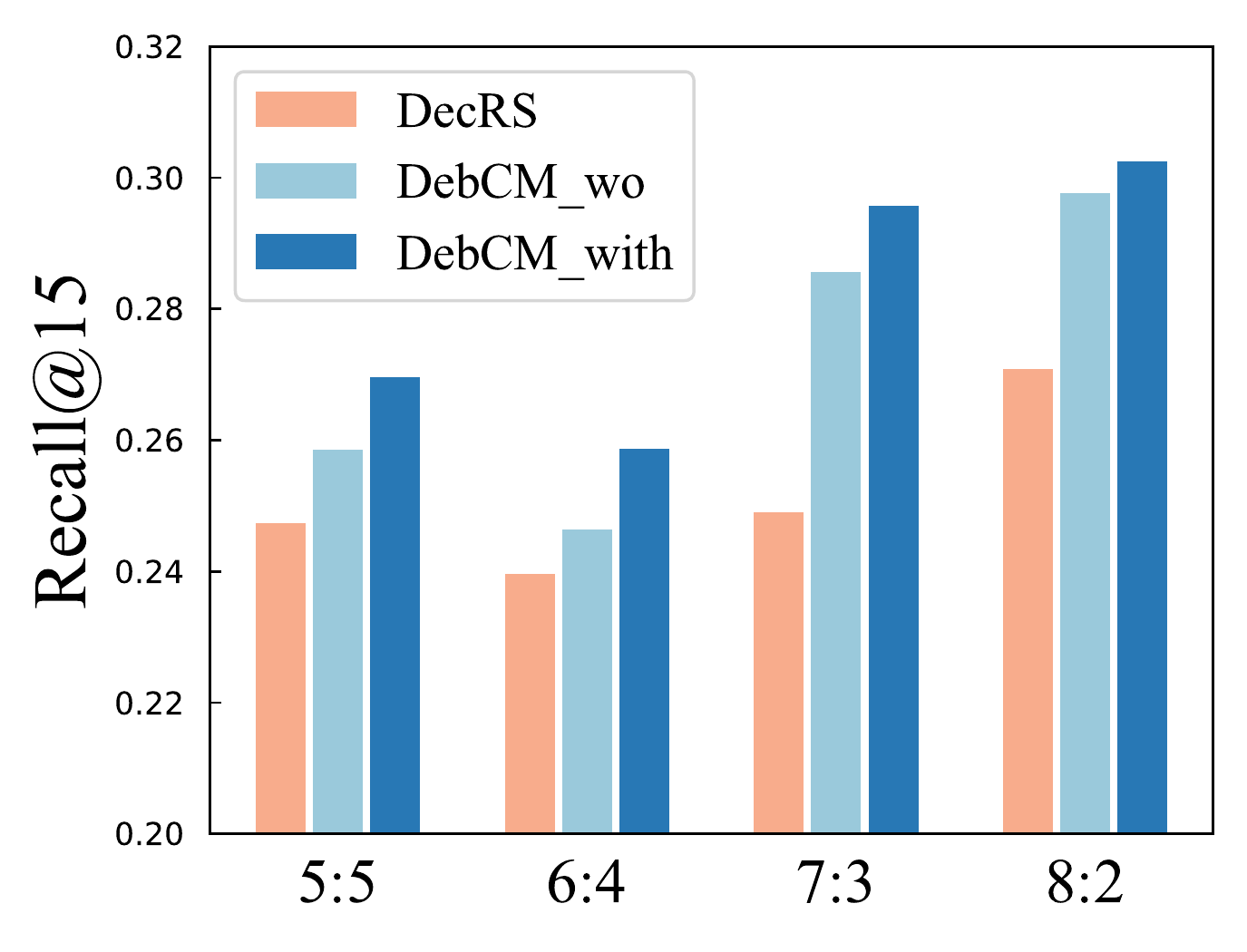}
}
\subfigure[NDCG@15]{
\centering
\includegraphics[scale=0.29]{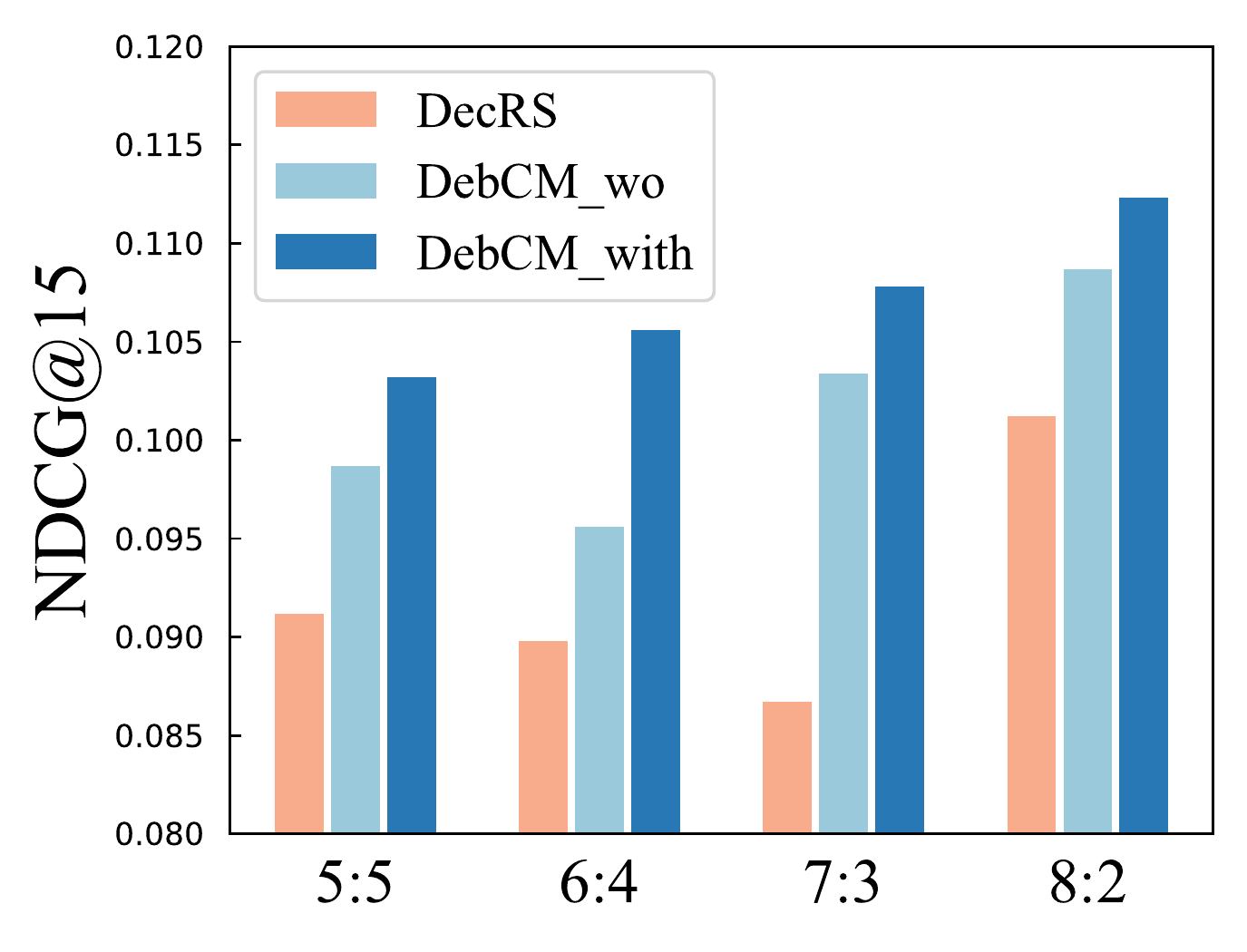}
}

\centering
\subfigure[Recall@25]{
\centering
\includegraphics[scale=0.29]{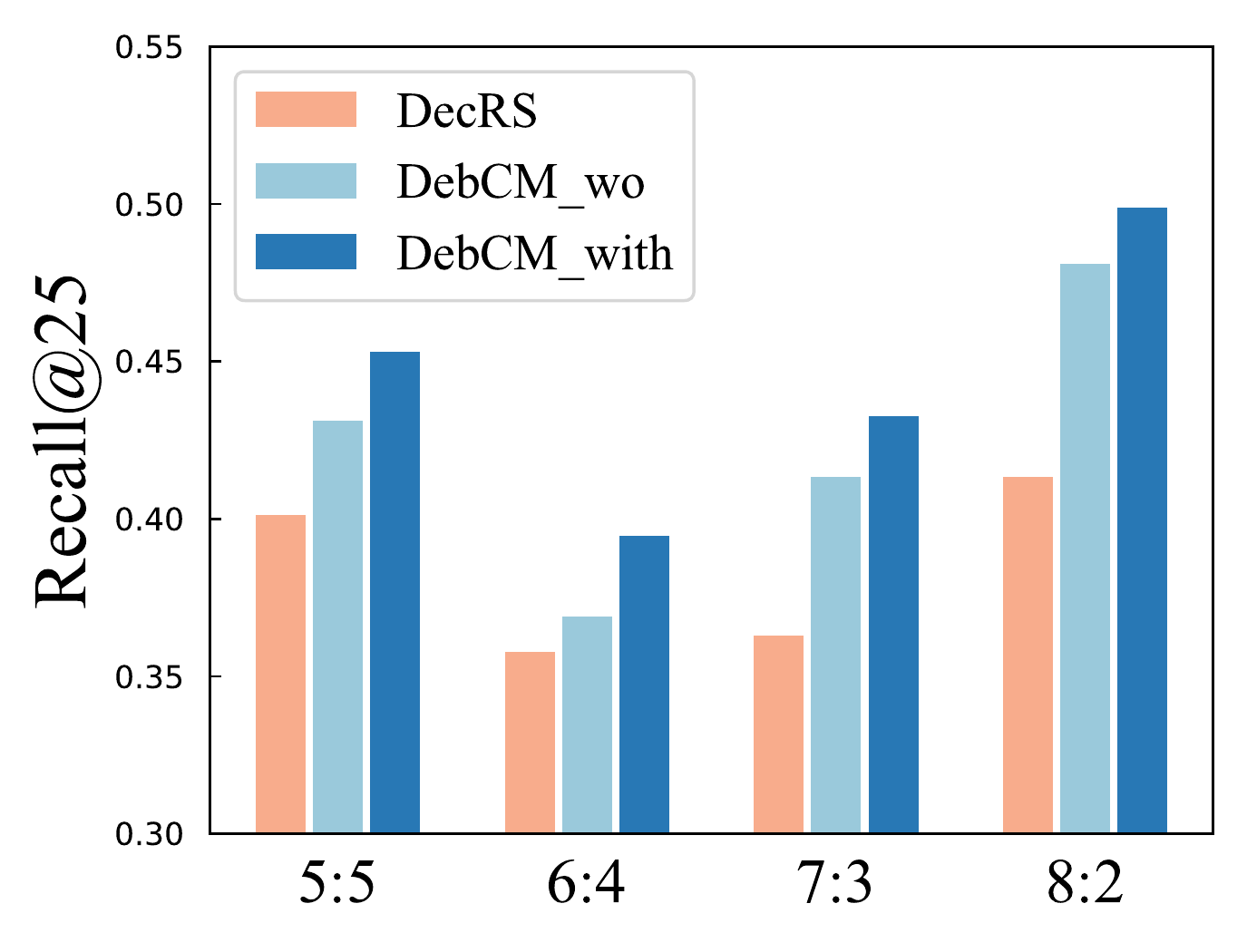}
}
\subfigure[NDCG@25]{
\centering
\includegraphics[scale=0.29]{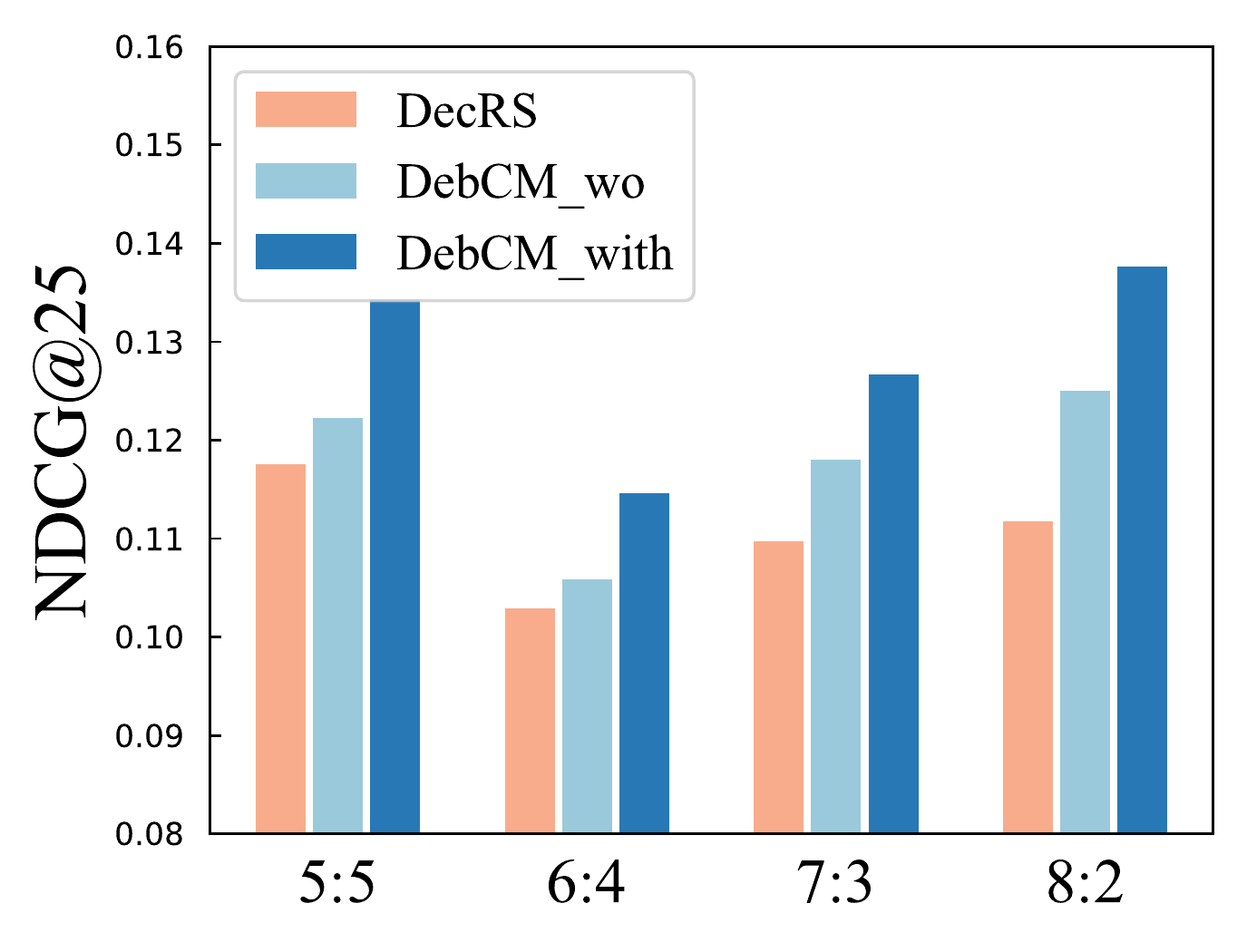}
}
\caption{Comparisons between DebCM and DecRS on TT-150k-genre dataset with varied test set intervention strength.}
\label{FIG:intervention_strength}
\end{figure}

\begin{figure}
\centering
\subfigure[Recall]{
\centering
\includegraphics[scale=0.29]{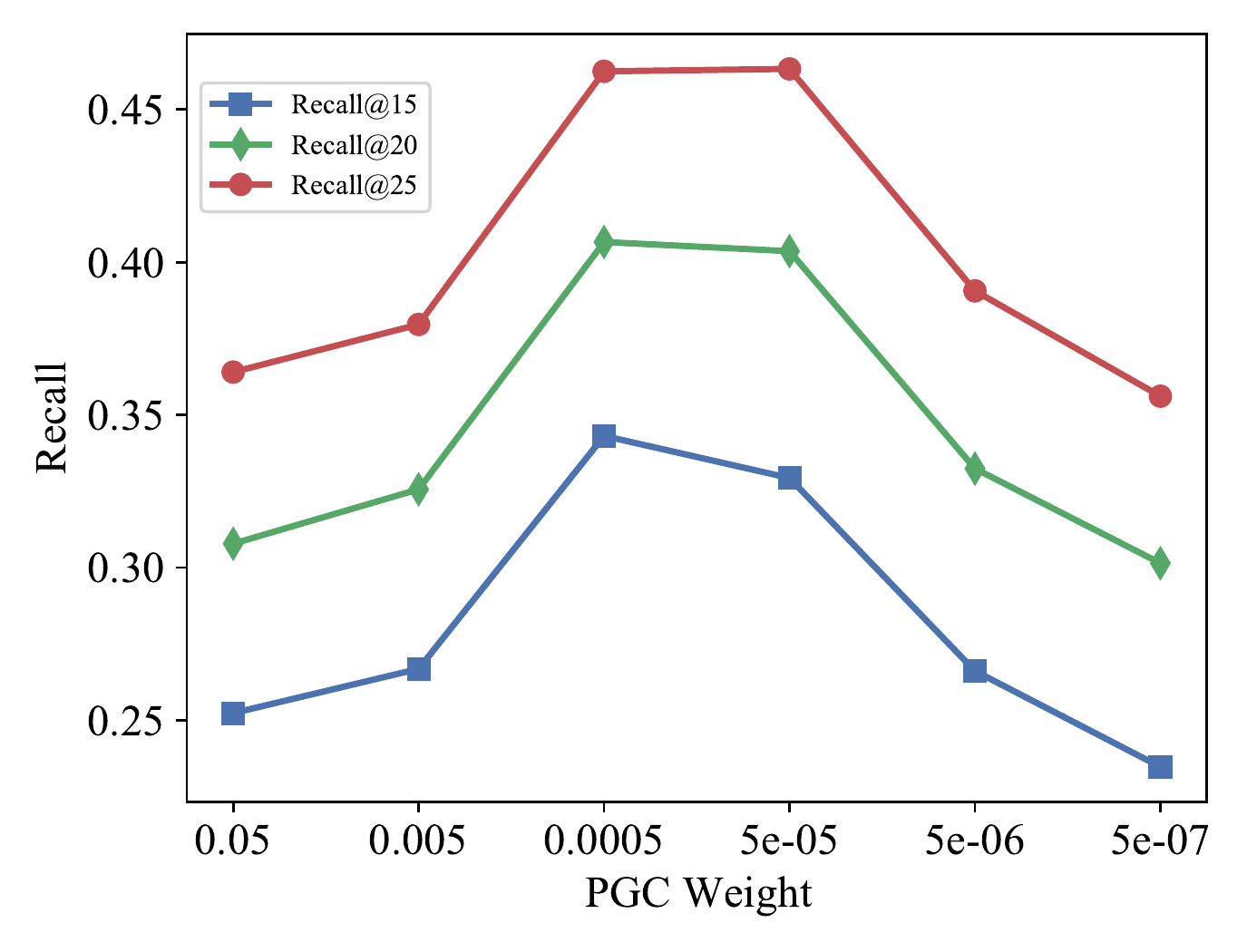}
}
\subfigure[NDCG]{
\centering
\includegraphics[scale=0.29]{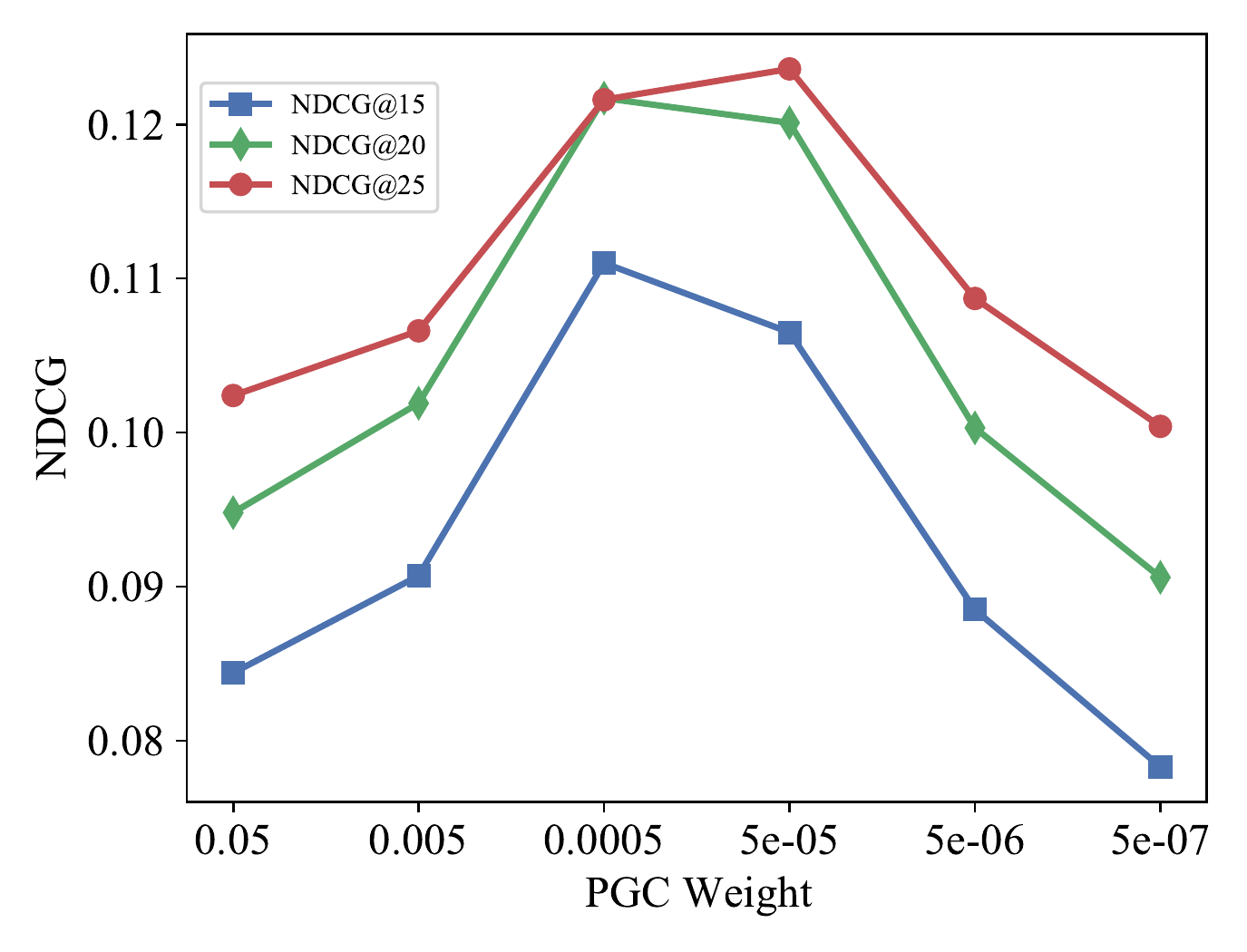}
}
\caption{Comparisons of different weights of the teacher loss.}
\label{FIG:weight_of_teacher}
\end{figure}


\begin{figure*}[h]
\centering
\includegraphics[scale=1.3]{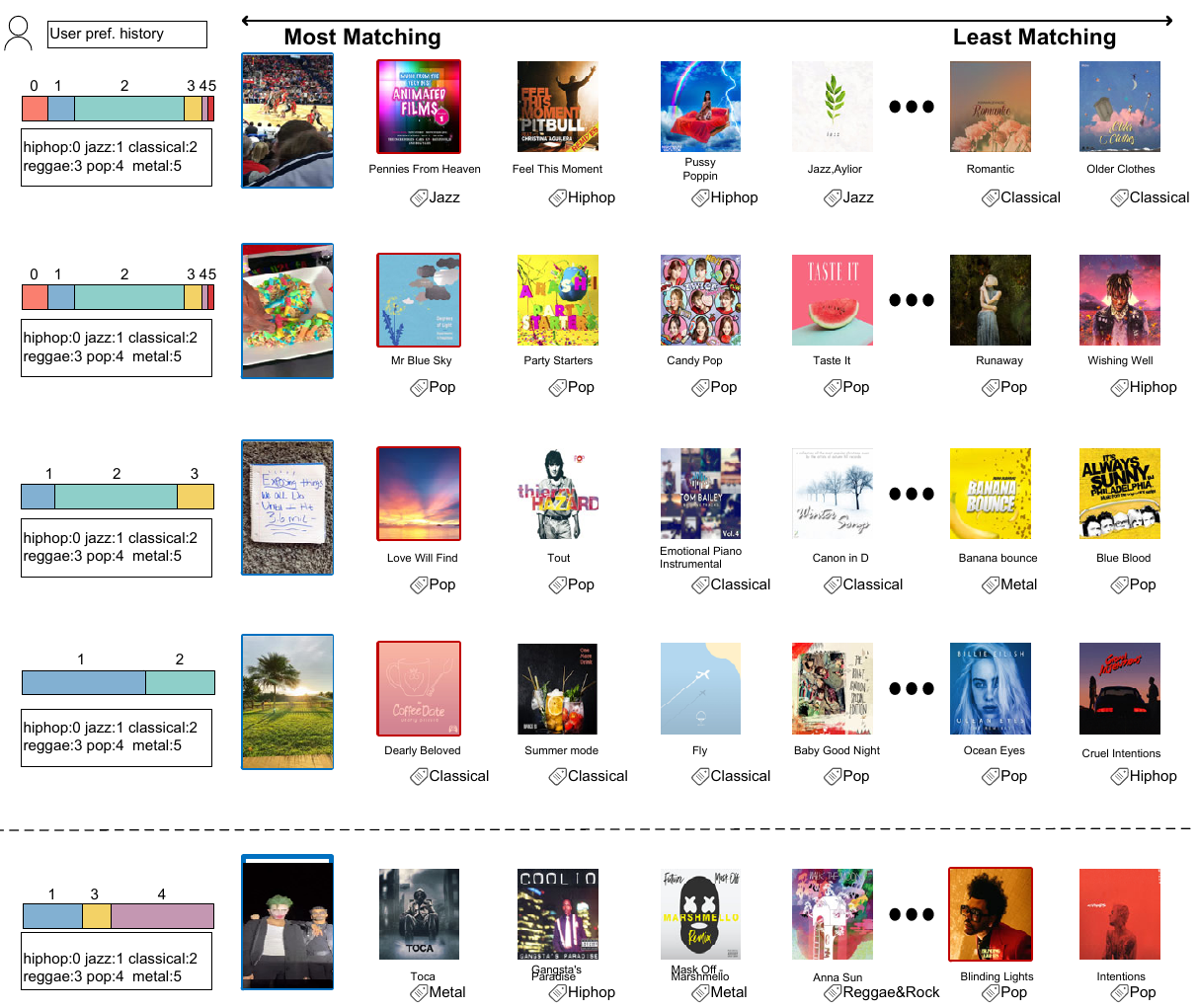}
\caption{Case study of DebCM for micro-video background music recommendation by visualizing some examples of videos where user interest shifts happened in the test set compared to the training set. The left column represents user historical preference among music genres, right columns depict the retrieved music clips in a most-to-least matching manner of the query videos ranked by our model. (pictures circled in blue are representative frames of the query videos and pictures circled in red are covers of uploader-selected ground-truth of music clips.)}
\label{FIG:visualization}
\end{figure*}

\textbf{\textit{Causal intervention on the dataset for deconfounder evaluation:}}
To evaluate the deconfounder ability of our model using backdoor adjustment, causal intervention is utilized to control music genre distributions in the training and test set. Specifically, to conduct an intervention on the TT-150k-genre dataset, the genre distribution in training and test sets for ``hiphop" and ``jazz" are set to be different, which are two common music genres the majority of users like both. For keeping the $X:(1-X)$ ($X \in range(0.1, 0.9)$) distribution of two music genres in the training set and the 8:1:1 ratio of training, validation, and test set, we calculate the number of music of genre ``hiphop" and genre ``jazz" in the training set, and then the ratio of two music genre in validation and test set approximates to $(1-X):X$, which exists a huge difference. We exhibit 4 different splits with $X$ set to be 0.8, 0.7, 0.6, and 0.5, where $X$ with a value close to extremity 1 demonstrates a stronger intervention imposed upon the dataset.
Moreover, under such a setting, we split the training and test set with music clips, where the music in the training set and test set are mutually exclusive. We follow the setting of ``strong generalization” in \cite{jy} where stratified sampling according to music popularity is utilized.

We explore the performance under different ratios of music genres of the training set and test set between our proposed model and the best-performing causal reasoning baseline DecRS to verify the robustness of our model to deviations caused by users' historical preferences in data. Moreover, we could analyze whether using the strategy of batch-level average to realize the approximation of backdoor adjustment proposed in this paper is better than the global mean in DecRS. Furthermore, we could investigate the robustness of our well-designed teacher-student network to different deviation degrees.

As seen from the experimental results in Fig. \ref{FIG:intervention_strength}, our model achieves the best performance under different music genre ratios of the test set. The comparison between DebCM and DecRS without PGC data demonstrates that the batch-level average strategy proposed in this paper introduces more randomness and conforms to the unbiased setting of Monte Carlo sampling, so as to achieve better results and show good performance for different degrees of data deviation.  Furthermore, we introduce PGC data for knowledge distillation to guide students' unbiased learning of the network, which further improves the model's debiasing effect and achieves the best performance under different deviation ratios.

\subsubsection{The Effectiveness of PGC Knowledge Distillation}

The performance of PGC data-induced knowledge distillation is investigated in this section, which is one of the main designs. Specifically, we explore the impact of different weights of the PGC network to explore the control weight of distillation.
We explore the influence of different weights of the teacher network on the training of the student network.  Specifically, we control the constraints of the teacher to the student network training, and the weight of PGC loss among \{5e-2, 5e-3, 5e-4, 5e-5, 5e-6, 5e-7\}.  With the increase of weight, the influence of the teacher network increase, thus more guidance of hidden variables of student network to be near to teacher network, so as to better remove the influence of user's personal preference effect. From the results in Fig. \ref{FIG:weight_of_teacher}, we observe that the performance rises first and then descends with the weight of the loss of teacher network increases.
It suggests that learning of hidden variables in the student network could be guided and constrained by hidden variables inferred by the teacher network, which removes the selection bias of the student network by guiding the uploader's personal matching with professional matching. However, excess constraints will limit the data fitting of the student network on its own UGC data, thereby degrading the performance of the network.  Therefore, we should choose an appropriate constraint weight to reasonably use PGC data to train the appropriate teacher network and give appropriate control to achieve the best constraint and guidance for the student network.

\subsection{Visualization}

This section visualizes videos selected from test sets and displays the sorted results obtained by our model, with the best match on the left and the least match on the right. We also show the historical preference among music genres of the video's uploader to see the performance of our model when user interest drift happens.   As can be seen from Fig. \ref{FIG:visualization}, our model can effectively reduce the effect of bias amplification of traditional recommendation systems for users whose historical preferences are concentrated in a certain genre. Instead, it recommends some background music that is more compatible with the video in the test set, which is also matched by users with interest shifts.  For example, in the first video, the uploader preferred classical music in history, while the uploaded video belongs to sports videos with more intense and high rhythm, which is not suitable for classical music. Therefore, this model recommends more lively and bright Jazz music and some intense Hip-hop music.  This makes matching more reasonable, which is achieved by deconfounding for user historical preference and learning good matches from PGC data.  However, the video below the dotted line is a bad case. The video's uploader used to be more inclined to pop music, while the model learned that the video is closer to reggae and Hip-hop styles, so it recommended such songs. However, the uploader still sticks to its preference and chooses pop music.  This also demonstrates the two considerations of uploaders when choosing background music: the user's own preference and the best-matching background music of the video.  How to balance these two aspects is a problem worthy of further study.  

\section{CONCLUSIONS}  \label{section6}

In this work, we have proposed a Teacher-Student network where the influence of user historical preference, as a confounder, is alleviated in the student network based on causal intervention. Specifically, the backdoor adjustment strategy is well approximated, where
uploader-induced confounding bias can be addressed
by substituting batch-level average uploader for the true uploader with personal bias. In addition, the matching of PGC data is captured by a teacher network which is capable of guiding the matching of the UGC student network by KL-based knowledge transfer. To evaluate the debiased ability of our proposed model, we extend the background music recommendation dataset TT-150k with music genre labels where an intervened evaluation strategy is introduced accordingly. Extensive experiments demonstrate
that DebCM is more robust to uploaders’ selection bias than
state-of-the-art strategies.

\bibliography{ms}

\end{document}